\documentclass[condensedmatter,article,moreauthors,dvi2pdf,10pt,a4paper]{mdpi}

\firstpage{1} 
\history{Received: 30 November 2017; Accepted: 30 January 2018; Published: date}

\usepackage{multirow}

\preto{\abstractkeywords}{\nolinenumbers}

\usepackage{amssymb}
\usepackage{amsmath}
\usepackage{amscd}
\usepackage{latexsym}

\usepackage{graphicx}

\newcommand{\be}{\begin{equation}}
\newcommand{\ee}{\end{equation}}
\newcommand{\Dlt}{\Delta}
\newcommand{\dlt}{\delta}

\newcommand{\vp}{\varphi}
\newcommand{\ep}{\varepsilon}
\newcommand{\al}{\alpha}
\newcommand{\bt}{\beta}

\newcommand{\gm}{\gamma}
\newcommand{\om}{\omega}

\newcommand{\bB}{{\bf B}}
\newcommand{\bS}{{\bf S}}
\newcommand{\bfe}{{\bf e}}

\newcommand{\rgl}{\rangle}
\newcommand{\lgl}{\langle}

\Title{Effects of Symmetry Breaking in Resonance Phenomena}

\Author{Vyacheslav I. Yukalov $^{1,2,}$*$^{,\dagger}$ and 
Elizaveta P. Yukalova $^{3,\dagger}$}
\AuthorNames{Vyacheslav I. Yukalov and Elizaveta P. Yukalova}

\address {%
$^{1}$ \quad Bogolubov Laboratory of Theoretical Physics,
Joint Institute for Nuclear Research, Dubna 141980, Russia \\
$^{2}$ \quad Instituto de Fisica de S\~ao Calros, Universidade de S\~ao Paulo, 
CP 369,  S\~ao Carlos 13560-970, S\~ao Paulo, Brazil \\
$^{3}$ \quad Laboratory of Information Technologies, 
Joint Institute for Nuclear Research, Dubna 141980, Russia; yukalova@theor.jinr.ru}

\corres{Correspondence: yukalov@theor.jinr.ru; Tel.: +7-496-216-3947}

\firstnote{These authors contributed equally to this work.}

\abstract{We show that resonance phenomena can be treated as nonequilibrium phase
transitions. Resonance phenomena, similar to equilibrium phase transitions, are 
accompanied by some kind of symmetry breaking and can be characterized by order 
parameters. This is demonstrated for spin-wave resonance, helicon resonance, and 
spin-reversal resonance.}


\begin{document}

\section{Introduction}

The great majority of phase transitions are characterized by spontaneous symmetry
breaking and can be described by the qualitative changes in the order parameter
behavior associated with long-range order \cite{Landau_1,Bogolubov_2}. This 
concerns both first-order as well as second-order phase transitions. There are 
also so-called topological phase transitions \cite{Fradkin_3} that are not
necessarily accompanied by symmetry breaking, but exhibit the changes in the
behavior of correlation functions and of reduced density matrices, connected with
a kind of quasi-long-range or mid-range order \cite{Coleman_4}. In all cases,
even when order parameters cannot be defined, different phases can be classified
by order indices of density matrices \cite{Coleman_5,Coleman_6,Coleman_7,Coleman_8}
quantifying all types of order, be they long-range or mid-range.

Phase transitions between equilibrium states of matter are induced by the variation in
thermodynamic parameters or static external fields. Similarly, the appearance of new
properties in a nonequilibrium system can be induced by alternating fields, especially
when some resonances occur.

In the present paper, we advocate the point of view that resonance phenomena can be
treated as nonequilibrium phase transitions. As in the case of equilibrium phase
transitions, resonance phenomena are accompanied by qualitative changes in their
macroscopic properties, which makes it possible to introduce related order parameters.
In many cases, resonance phenomena exhibit a kind of symmetry breaking. We
illustrate these properties by considering several resonances for which we define the
related order parameters and show that this kind of symmetry can become broken. As
examples, we consider spin-wave resonance, helicon resonance, and spin-reversal
resonance.

Let us emphasize that we do not claim that resonance phenomena are exactly the same 
as equilibrium phase transitions. This is evidently not so, since resonance phenomena 
describe nonequilibrium processes. However, we show that both these phenomena can 
share two, probably the most important, properties: First, there can occur some symmetry 
breaking in a region around the resonance. Second, it is possible to introduce order 
parameters distinguishing between qualitatively different states of the considered system.     
These similarities justify the comparison between resonance phenomena and phase transitions. 

We do not prescribe in advance what would be the order of transition in particular resonance
phenomena. As always, this is defined by the behavior of the related order parameters. The 
realistic description of resonance phenomena is usually dealt with finite systems, since, to
realize a resonance, one always needs an alternating external field, created outside of and 
acting on the system. For finite systems, the problem of thermodynamic limit does not arise 
at all. Dealing with finite systems, it is natural to expect that the related resonance 
transitions will be of continuous type, analogous to transformations in finite equilibrium 
systems. In the majority of cases, transitions caused by resonances are in fact crossover 
transitions. 

An exact resonance, as such, happens at a single point of a varied parameter, usually of
frequency. However, this does not mean that the transition is localized at that single point. 
As we shall see from the examples below, there is a region around the point of the exact 
resonance where new properties, compared to those in the regions far from the     
resonance point, arise. Such a region, which can be called the {\it resonance region}, reminds us of the
critical region in equilibrium phase transitions. 

Again let us stress that resonance phenomena are nonequilibrium, and symmetry breaking,
generally, drives systems to nonuniform states. The most convenient way of describing such
systems is by considering the dynamics of observable quantities that are defined through the
corresponding statistical averages. All examples, we consider below, are based on exactly
this approach of studying the dynamic behavior of observable quantities. The dynamical 
equations in all cases are derived from related microscopic theories. Of course, it 
would be absolutely unreasonable to reproduce all these rather complicated derivations in 
the present paper. This could take hundreds of pages, especially for realistic problems we 
deal with. Instead, it is sufficient in the present paper to give the appropriate references where the reader 
can find all details.

Throughout the paper, we set the Planck constant to unity.

\section{Spin-Helicon Waves}

First, we consider resonances that occur in a paramagnetic metal subject to the irradiation
of an external electromagnetic field. The metal is assumed to have the geometry of a plate
in the region $0 < z < L$. There is an external static magnetic field along the $z$ axis,
\be
\label{1}
 \bB_0 = B_0 \bfe_z \; 
\ee
and perpendicularly to its surface the metal is irradiated by an alternating electromagnetic
field of frequency $\omega$ much lower than the plasma frequency,
\be
\label{2}
 \om \ll \om_p \qquad \left ( \om_p^2 \equiv 4\pi\; \frac{\rho e_0^2}{m} \right ) \; 
\ee
where $\rho\sim 10^{22}$ cm$^{-3}$, $e_0$, and $m$ are the electron density, charge,
and mass, respectively. The plasma frequency is $\omega_p \sim 10^{15}$ s$^{-1}$.
The static paramagnetic susceptibility in paramagnetic metals is small, so~that
\be
\label{3}
  4\pi\chi \ll 1 \; .
\ee

Usually, the susceptibility is of order $\chi \sim 10^{-6}$.

The excitation of waves inside a metallic plate is achieved in the best way by circularly
polarized electromagnetic waves \cite{Kaner_9,Falko_10}, because of which we consider
the electromagnetic fields and magnetic moment in the form of the combinations
\be
\label{4}
 H = H_x - i H_y \; , \qquad E = E_x - i E_y \; , \qquad M = M_x - i M_y \;  .
\ee

The temporal behavior is described by $\exp(-i \omega t)$.

The coupled Maxwell--Bloch equations for linear field deviations in a paramagnetic
metal  with weak dispersion and isotropic Fermi surface can be written
\cite{Silin_11,Platzmann_12,Yukalov_13,Yukalov_14,Yukalov_15} as
$$
\frac{dH}{dz} \; - \; \ep k E = 0 \; , \qquad 
\frac{dE}{dz} \; + \;  k (H + 4\pi M ) = 0 \;
$$
\be
\label{5}
\left (D\; \frac{d^2}{dz^2} \; - \; \om_s + i \nu_s \right ) ( M - \chi H ) + \om M = 0 \; .
\ee

Here, $k \equiv \omega/c$, the effective dielectric permeability is
\be
\label{6}
 \ep = - \; \frac{\om_p^2}{\om(\om-\om_s+ i\nu_0) } \; 
\ee
and the diffusion coefficient reads as
\be
\label{7}
 D =  \frac{v_F^2(1+\bt_0)(1+\bt_1)}{3(\om-\om_0+ i\nu_0) } \;  
\ee
where
\be
\label{8}
\om_s = \frac{e_0B_0}{mc}
\ee
is the Larmor spin frequency, and the cyclotron frequency
\be
\label{9}
\om_0 = \frac{1+\bt_1}{1+\bt_0} \; \om_s
\ee
is renormalized by the Landau Fermi-liquid interaction parameters $\beta_0$ and $\beta_1$.
The attenuations
\be
\label{10}
\nu_s = \frac{1+\bt_0}{\tau_s} \; , \qquad \nu_0 = \frac{1+\bt_1}{\tau_0} 
\ee
are defined by the times of momentum, $\tau_0 \sim 10^{-9}$ s, and spin, $\tau_s \sim 10^{-6}$ s,
relaxations. The Fermi velocity of conduction electrons is $v_F \sim 10^8$ cm s$^{-1}$.
Equations (\ref{5}) 
describe the coupled spin-helicon waves with the dispersion relation
\be
\label{11}
 q^2 = \ep \mu(q,\om) \;  
\ee
where the effective magnetic permeability is
\be
\label{12}
\mu(q,\om) = \frac{\om-\om_s+i\nu_s-Dq^2}{\om+(1-4\pi\chi)(-\om_s+i\nu_s-Dq^2)} \;   .
\ee

This gives us two solutions for the characteristic wave vectors, associated with the
spin waves, $k_s$, and helicon waves, $k_h$. Taking into account the smallness of
the static paramagnetic susceptibility, we~can write
\be
\label{13}
k_s^2 = \frac{\om-\om_s+i\nu_s}{D} \; , \qquad k_h^2 = \ep k^2 \qquad (4\pi\chi \ll 1 )
\ee
to zero order in $\chi$, and
$$
k_s^2 = \frac{\om-\om_s+i\nu_s}{D} \; \left ( 1 + 
\frac{4\pi\chi\om}{\om-\om_s+i\nu_s-D\ep k^2} \right ) \; 
$$
\be
\label{14}
k_h^2 = \ep k^2  \left ( 1 + 4\pi\chi\;
\frac{-\om_s + i\nu_s - D\ep k^2}{\om-\om_s+i\nu_s-D\ep k^2} \right ) \; 
\ee
to first order in $\chi$.

The incident and reflected fields are plane waves, with the magnetic components
\be
\label{15}
H_0(z) = H_0 e^{ikz} \; , \qquad H_1(z) = H_1 e^{-ikz} \qquad ( z \leq 0 ) \; ,
\ee
which defines the total magnetic field $H_0(z) + H_1(z)$. Respectively, the electric
components yield the electric field
\be
\label{16}
 E_0(z) + E_1(z) = i [ H_0(z) - H_1(z) ] \qquad ( z\leq 0) \;  .
\ee

Inside the metallic plate, the magnetic field consists of four parts, including two running
waves and two waves reflected from the second surface of the plate,
\be
\label{17}
 H(z) = H_2 e^{ik_sz} +  H_3 e^{-ik_sz} + H_4 e^{ik_hz} + H_5 e^{-ik_hz} 
\qquad ( 0 \leq z \leq L) \; .
\ee

The field transmitted through the second surface is
\be
\label{18}
 H_6(z) = H_6 e^{ikz} \; , \qquad E_6(z) = i H_6(z) \qquad ( z \geq L ) \; .
\ee

Equations (\ref{5}) are to be complimented by the boundary conditions. For magnetic
and electric fields, there are the standard continuity conditions for their tangential
components on each of the surfaces. The spatial structure of the metallic surface
influences the spatial distribution function of conducting electrons \cite{Askerov_16}.
This can result in the appearance of a magnetic anisotropy on the surface. Several
boundary conditions for the magnetization have been studied
\cite{Walker_17,Janossy_18,Menard_19,Flesner_20,Silsbee_21,Graham_22}.
A simple boundary condition for the magnetization was proposed by Dyson \cite{Dyson_23},
which reads as
\be
\label{19}
 \frac{dM}{d{\bf n}} + \zeta M = 0 \qquad ( z = 0, L ) \; 
\ee
where the first term implies the normal derivative at the boundary and $\zeta$ is a surface
anisotropy parameter connected to the probability of a spin flip when scattering at the
surface. The Dyson boundary condition has been used in several papers
\cite{Janossy_18,Yukalov_24,Yukalov_25,Yukalov_26}. Experiments have not been able to determine
between the preferred type of condition \cite{Magno_27,Vander_28}. Hence, without the
loss of generality, we can employ the Dyson condition (\ref{19}).

The six boundary conditions at two surfaces, for the fields and for the magnetic moment,
define~all six amplitudes $H_i$, with $i = 1,2,3,4,5,6$ as functions of the incident-field
amplitude $H_0$.

A convenient observable quantity is the transparency coefficient
\be
\label{20}
C_T \equiv \left | \; \frac{H_6}{H_0} \; \right |^2 
\ee
showing how the incident electromagnetic field is transmitted through the metallic plate.

\section{Spin-Wave Resonance}

When the frequency $\omega$ of the incident field is close to the spin frequency
$\omega_s$, the helicon amplitudes $H_4$ and $H_5$ are small, as compared to the
spin-wave amplitudes $H_2$ and $H_3$, and the spin wave forms a standing wave,
so that the magnetic field inside the plate becomes practically periodic, slightly~perturbed by attenuations. Respectively, the magnetic moment is also practically
periodic:
\be
\label{21}
 M(z) \; \propto \; \cos(k_s z) \;  .
\ee

This looks like a kind of magnetic crystallization of spin-wave collective excitations,
whereby the system becomes spatially periodic if a small attenuation is neglected.

For typical paramagnetic metals, the spin frequency is $\omega_s \sim 10^{11}$ s$^{-1}$,
the wave vectors are $k_s \sim 10^2$ cm$^{-1}$ and $k_h \sim 10^6$ cm$^{-1}$, and the
magnetic anisotropy parameter is $\zeta \sim 10^3$cm$^{-1}$. The plate width is typically
$L \sim 10^{-2}$ cm. Hence the inequalities
\be
\label{22}
|\; k_s \; | \ll \zeta \ll |\; k_h \; | \; \qquad |\; k_h L \; | \gg 1
\ee
are valid, which will be used in what follows. Employing these inequalities, we
obtain the transparency coefficient
\be
\label{23}
 C_T = \frac{64\pi^2\chi^2\om^2|\;k_s \;|^2}{c^2\zeta^6 |\; \sin(k_sL) \; |^2} \qquad
( \om \sim \om_s) \; .
\ee
In the denominator,
$$
|\; \sin(k_sL) \; |^2 = \sin^2({\rm Re}k_s L ) + \sinh^2({\rm Im}k_sL ) \; 
$$
where
$$
{\rm Re}k_s = |\;k_s\;| \cos\vp \; ,  \qquad  {\rm Im}k_s = |\;k_s\;| \sin\vp \; ,
$$
and $\varphi$ is the argument of $k_s$. Thus, we have
\be
\label{24}
C_T \; \propto \; \frac{\om^2|\;k_s\;|^2 L^2}
{\sin^2(|\;k_sL\;| \cos\vp) + \sinh^2(|\;k_sL\;|\sin\vp)} \; 
\ee
where
\be
\label{25}
|\;k_sL\;| = \frac{\sqrt{3}}{v_F} \; \left \{ 
\frac{[\; (\om-\om_s)^2+\nu_s^2\;][\;(\om-\om_0)^2+\nu_0^2\;]}{(1+\bt_0)^2(1+\bt_1)^2} 
\right \}^{1/4}
\ee
and the phase is
\be
\label{26}
 \vp = \frac{1}{2} \; \arctan \left [ \;
 \frac{\nu_0(\om-\om_s)+\nu_s(\om-\om_0)}{(\om-\om_s)(\om-\om_0) -\nu_0\nu_s} \;
\right ] \;  .
\ee

The spin-wave resonance occurs under the condition
\be
\label{27}
{\rm Re}
 k_s L = |\;k_s L\;| \cos\vp = \pi n \qquad ( n = 1,2,\ldots ) \; .
\ee

To simplify the formulas, we can take into account that the Fermi-liquid interaction
parameters $\beta_0$ and $\beta_1$ are small and that dimensionless attenuations
\be
\label{28}
\nu_1 \equiv \frac{\nu_0}{\om_s} \; , \qquad  \nu_2 \equiv \frac{\nu_s}{\om_s} \;  .
\ee

In typical paramagnetic metals, $\nu_0 \sim 10^9$ s$^{-1}$, $\nu_s \sim 10^6$s$^{-1}$, and
$\omega_s \sim 10^{11}$ s$^{-1}$. Therefore, $\nu_1 \sim 10^{-2}$ and $\nu_2 \sim 10^{-5}$.
In view of these small parameters, we have
$$
|\;k_s\;| \simeq \frac{\sqrt{3}}{v_F} \; ( \om - \om_s ) \;  .
$$

Thus, the resonance condition (\ref{27}) yields the spin-resonance frequencies
\be
\label{29}
\om_n = \om_s \left ( 1 + \frac{An}{\cos\vp} \right ) \qquad ( n = 1,2,\ldots ) \;  
\ee
in which
\be
\label{30}
  A \equiv \frac{\pi v_F}{\sqrt{3} \; L\om_s} \; .
\ee

For the typical values $v_F \sim 10^8$ cm s$^{-1}$ and $L \sim 10^{-2}$ cm, the parameter
$A \sim 0.1$ and $\varphi \sim 10^{-2}$. Hence,
\be
\label{31}
 \vp \simeq \frac{1}{2} \; \arctan \left ( \frac{\nu_1+\nu_2}{An} \right ) \;  .
\ee

In what follows, we consider the first resonance, with $n = 1$, whose frequency is
\be
\label{32}
  \om_1 = \om_s ( 1 + A ) \;
\ee
where we take into account that, because of the smallness of $\varphi$, 
$\cos \varphi \approx 1$.

Introducing the relative detuning
\be
\label{33}
 \dlt \equiv \frac{\om-\om_1}{\om_1} \; 
\ee

we can write
$$
 |\;k_s L\;| = \pi ( 1 + b\dlt ) \qquad 
\left ( b \equiv \frac{1+A}{A} \right ) \; .
$$

The occurrence of spin-wave resonance manifests itself by the appearance of the
well observable property of the transparency of the metallic plate with respect to
the penetration of electromagnetic waves. The order parameter can be defined as
the normalized transparency coefficient
\be
\label{34}
\eta \equiv \frac{C_T(\dlt)}{C_T(0)} \;   .
\ee

For the latter, we obtain
\be
\label{35}
 \eta = \frac{(1+\dlt)^2(1+b\dlt)^2\sinh^2(\pi\sin\vp)}{\sin^2[\pi(1+b\dlt)\cos\vp]
+ \sinh^2[\pi(1+b\dlt)\sin\vp] } \; .
\ee

\begin{figure}[H]
\centering
\includegraphics[width=6cm]{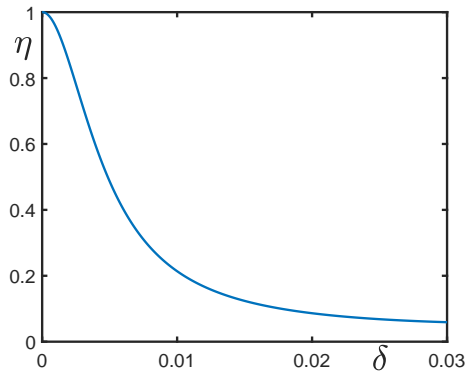}
\caption{Order parameter (\ref{35}) characterizing the plate transparency under a
spin-wave resonance, with the parameters $A = 0.1$, $\nu_1 = 10^{-2}$, and
$\nu_2 = 10^{-5}$.}
 \label{f1}
\end{figure}

The order parameter (Equation (\ref{35})), as a function of the relative detuning, is shown in
Figure \ref{f1}. In the vicinity of the resonance frequency, where the detuning is close to zero,
the order parameter is close to one. In addition, it diminishes with increasing the detuning.

The state at small detuning is qualitatively different from the state far from $\delta = 0$.
The sample under spin-wave resonance becomes periodic due to the developed
standing wave (\ref{21}), and the plate exhibits the macroscopic property of transparency.
These features disappear outside of the resonance. Such behavior reminds us of a
phase transition.

\section{Helicon Resonance}

At a frequency $\omega$ much lower than the spin frequency $\omega_s$,
\be
\label{36}
 \om \ll \om_s \; 
\ee
spin waves strongly attenuate, but helicon waves can persist \cite{Petrashov_29,Kotelnikov_30}.
At such frequencies, the transparency coefficient (\ref{20}) becomes
\be
\label{37}
C_T = \frac{4\om\om_s}{\om_p^2|\;\sin(k_hL)\;|^2} \;  
\ee
where
\be
\label{38}
 k_h = \sqrt{\ep} \; \frac{\om}{c} \; , \qquad 
\ep = \frac{\om_p^2(\om_s+i\nu_0)}{\om(\om_s^2+\nu_0^2)} \;  .
\ee

The real and imaginary parts of the helicon wave vector are
\be
\label{39}
{\rm Re}k_h = \frac{\om_p}{c} \; \sqrt{ \frac{\om}{\om_s}} \; \cos\vp \;  , 
\qquad
{\rm Im}k_h = \frac{\om_p}{c} \; \sqrt{ \frac{\om}{\om_s}} \; \sin\vp \;
\ee
with the argument
\be
\label{40}
 \vp = \frac{1}{2}\; \arctan\; \frac{\nu_0}{\om_s} \;  .
\ee

From the transparency coefficient
\be
\label{41}
C_T = \frac{4\om\om_s}{\om_p^2[\sin^2({\rm Re}k_hL)+\sinh^2({\rm Im}k_h L)]} 
\ee
it follows that the helicon resonance happens when
\be
\label{42}
 {\rm Re}k_h L = \pi n \qquad ( n = 1,2,\ldots ) \; .
\ee

The helicon resonance frequencies, keeping in mind that $\nu_0/ \omega_s \sim 10^{-2}$,
is read as
\be
\label{43}
 \om_n = \frac{\pi^2 c^2\om_s n^2}{\om_p^2 L^2 \cos^2\vp} \qquad  
( n = 1,2,\ldots ) \; .
\ee

We shall consider the first resonance with the frequency
\be
\label{44}
 \om_1 = \left ( \frac{\pi c}{\om_p L} \right )^2 \; \om_s \;  .
\ee

This frequency is of order $\omega_1 \sim 10^6$ s$^{-1}$.

The order parameter can again be defined as the normalized transparency coefficient
$$
 \eta \equiv \frac{C_T(\dlt)}{C_T(0)} \; , \qquad 
\dlt \equiv \frac{\om-\om_1}{\om_1} \;
$$
being a function of the relative detuning. Using the quantities
$$
{\rm Re}k_h L = \pi \; \sqrt{1+\dlt} \; , \qquad 
{\rm Im}k_h L =  \pi\; \sqrt{1+\dlt} \; \tan \vp 
$$
results in the order parameter
\be
\label{45}
 \eta = \frac{(1+\dlt)\sinh^2(\pi\tan\vp)}
{\sin^2(\pi\sqrt{1+\dlt})+\sinh^2(\pi\sqrt{1+\dlt}\tan\vp)}\;  .
\ee

\begin{figure}[H]
\centering
\includegraphics[width=6cm]{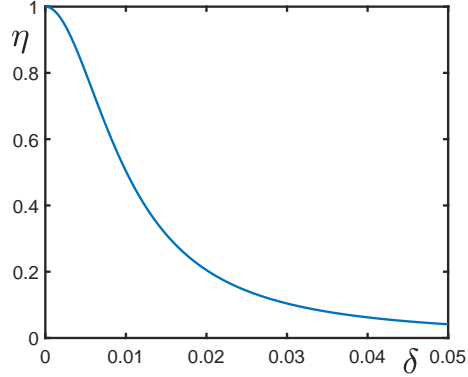}
\caption{Order parameter (\ref{45}) describing the plate transparency under
helicon resonance, with the parameter $\nu_1 = 10^{-2}$.}
\label{f2}
\end{figure}

The order parameter (Equation (\ref{45})) is shown in Figure \ref{f2}. The situation is similar to the case
of spin-wave resonance. At small detuning, a standing periodic
magnetic field develops, and the state is characterized by a large transparency. Outside of the resonance, the plate is not transparent.

In a similar way, we could describe other magnetic resonance phenomena, e.g.,
ferromagnetic resonance \cite{Akhiezer_31}. Some quantum mechanical scattering
problems, dealing with finite-width systems, also lead to equations exhibiting resonances
analogous to that considered above \cite{Zakhariev_32,Zakhariev_33,Zakhariev_34}.
For such problems, it is also possible to introduce order parameters as normalized
transparency coefficients.

\section{Spin-Rotation Symmetry}

Let us consider a lattice of $N$ lattice sites, with a spin operator ${\bf S}_j$ in the
$j$-th site, where $j = 1, 2, \ldots , N$. The system is placed in a magnetic field
${\bf B}_0 = B_0 {\bf e}_z$ along the $z$-axis. The Hamiltonian is
\be
\label{46}
 \hat H_0 = -\mu_0 B_0 
\sum_{j=1}^N S_j^z + \frac{1}{2} \sum_{i\neq j} \hat H_{ij} \;
\ee
in which $\mu_0 =-g_S \mu_B$, with $g_S$ being a $g$-factor and $\mu_B$, being
the Bohr magneton. The exchange spin~interactions
\be
\label{47}
\hat H_{ij} = - J_{ij} \left ( S_i^x S_j^x + S_i^y S_j^y \right ) - 
I_{ij} S_i^z S_j^z
\ee
correspond to the so-called $XXZ$ model.

The Hamiltonian is invariant with respect  to spin rotations around the $z$-axis. The
rotation operator~is
\be
\label{48}
 \hat R_z = \exp \left ( -i\vp S^z \right ) \;
\ee
where $\varphi$ is a rotation angle and
\be
\label{49}
S^z \equiv \sum_{j=1}^N S_j^z 
\ee
is the $z$-component of the total lattice spin. Because of the commutation relations
$$
 \left [ S^z , \; S_i^x S_j^x + S_i^y S_j^y \right ] = 
\left [ S^z , \; S_i^z S_j^z \right ] = 0 \;
$$
the $z$-component of the total spin is conserved:
$$
 \left [ S^z , \; \hat H_0 \right ] = 0 \;  .
$$

Therefore, the Hamiltonian is invariant under the rotation transformation,
\be
\label{50}
  \hat R_z^+ \hat H_0 \hat R_z = \hat H_0 \;
\ee
which implies the symmetry with respect to the spin rotation around the $z$-axis.
Thus, the Hamiltonian symmetry is $U(1)$.

Because of this symmetry, the transverse components of the average spin of
an equilibrium system are zero:
\be
\label{51}
 \lgl S_j^x \rgl =  \lgl S_j^y \rgl = 0 \; .
\ee

Respectively, the average values of the ladder operators
$$
S_j^\pm \equiv S_j^x \pm i S_j^y
$$
are also zero:
\be
\label{52}
  \lgl S_j^+ \rgl =  \lgl S_j^- \rgl = 0 \;   .
\ee

Suppose that, at the initial moment of time, the system is prepared as described above.
However, if the system is made nonequilibrium, the average spin can start changing,
breaking the $U(1)$ symmetry. This can also be accompanied by a reversal of the
total average spin.

A nonequilibrium spin system is characterized by the time-behavior of the following
quantities: the transition function
\be
\label{53}
  u \equiv \frac{1}{NS} \sum_{j=1}^N  \lgl S_j^- \rgl \;
\ee
the coherence intensity
\be
\label{54}
 w \equiv \frac{1}{N(N-1)S^2} \sum_{i\neq j}^N  \lgl S_i^+ S_j^- \rgl \;
\ee
and the average spin projection
\be
\label{55}
 s \equiv \frac{1}{NS} \sum_{j=1}^N  \lgl S_j^z \rgl \;  .
\ee

The details of temporal behavior of a nonequilibrium system depend on the type
of conditions transforming the system to a nonequilibrium state.

\section{Spin-Reversal Resonance}

The system, described in the previous section, then is connected to a resonator
electric circuit, and an additional transverse magnetic field $H$ starts acting on
the sample. Thus, the total magnetic field~becomes
\be
\label{56}
 \bB = B_0 \bfe_z + H \bfe_x \;  .
\ee

The important point is that this additional field $H$ is not just an external field,
but a feedback field created by the moving spins of the system. The equation
for the feedback field can be derived \cite{Yukalov_35,Yukalov_36,Yukalov_37}
from the Kirchhoff equation, yielding
\be
\label{57}
 \frac{dH}{dt} + 2\gm H + \om^2 \int_0^t H(t') \; dt' = - 4\pi\; \frac{dm_x}{dt} \;
\ee
where $\omega$ is the resonator natural frequency, $\gamma$ is the resonator
attenuation, and the electromotive force is caused by moving spins with the
magnetization density
\be
\label{58}
 m_x = \frac{\mu_0}{V_{res}} \sum_{j=1}^N \lgl S_j^x \rgl \;
\ee
with $V_{res}$ being the resonator coil volume.

Switching on the additional feedback field leads to the Hamiltonian
$$
 \hat H = \hat H_0 - \mu_0 H \sum_{j=1}^N  S_j^x \;  .
$$

Thus, the total Hamiltonian becomes
\be
\label{59}
 \hat H = -\mu_0 \sum_{j=1}^N \bB \cdot \bS_j \; + \;
\frac{1}{2} \sum_{i\neq j}^N \hat H_{ij} \;  .
\ee

It should be mentioned here that spins formed by electrons cause the negative magnetic moment
$\mu_0 < 0$. When~$B_0 > 0$, a positive value of the Zeeman
frequency results:
\be
\label{60}
 \om_0 = -\mu_0 B_0 > 0 \;  .
\ee

The coupling of the spin system to a resonator producing a feedback field defines
the feedback~attenuation
\be
\label{61}
 \gm_0 \equiv \pi \mu_0^2 S \; \frac{N}{V_{res}} \;  .
\ee

The effective coupling parameter, characterizing the interaction of the system with
the resonator,~is
\be
\label{62}
 g \equiv \frac{\gm_0 \om_0}{\gm\gm_2} \qquad ( \gm_2 \equiv \rho \mu_0^2 S ) \;
\ee
where $\rho$ is the spin density.

An efficient interaction between the system and the resonator can develop only when
the Zeeman frequency (Equation (\ref{60})) is close to the resonator natural frequency $\omega$ and
hence when the detuning
\be
\label{63}
\dlt \equiv \frac{\Dlt}{\om_0} = \frac{\om-\om_0}{\om_0}
\ee
is small. Additionally, all attenuations in the system need to be small compared to $\omega$
or $\omega_0$, and these attenuations  
include the resonator attenuation $\gamma$, feedback attenuation
$\gamma_0$, longitudinal attenuation $\gamma_1$, transverse attenuation $\gamma_2$,
and the spin-wave attenuation $\gamma_3$:
\be
\label{64}
  \frac{\gm}{\om} \ll 1 \; , \qquad \frac{\gm_0}{\om_0} \ll 1 \; , \qquad 
\frac{\gm_1}{\om_0} \ll 1 \; , \qquad \frac{\gm_2}{\om_0} \ll 1 \; , \qquad
\frac{\gm_3}{\om_0} \ll 1 \;  .
\ee

Finally, the relative anisotropy parameter
\be
\label{65}
  A \equiv \frac{S\Dlt J}{\om_0} \; , \qquad 
\Dlt J \equiv \frac{1}{N} \sum_{i\neq j}^N (I_{ij} - J_{ij} ) \;
\ee
should also be small so that the initial spin polarization is not blocked by
the anisotropy field and the latter does not create an essential dynamical shift
of the frequency
\be
\label{66}
 \om_s \equiv \om_0 ( 1 - As ) \;
\ee
thus producing a large effective detuning
\be
\label{67}
 \Dlt_s \equiv \om - \om_s = \Dlt + \om_0 A s \;  .
\ee

The existence of the above small parameters makes it possible to analyze the evolution
equations for the functional variables (Equations (\ref{53})--(\ref{55})) by the scale separation
approach \cite{Yukalov_38,Yukalov_39}, since the functional variable $u$ can be treated
as fast, and $w$ and $s$ treated as slow. In the frame of this approach, with the use
of the stochastic mean-field approximation, we come to the equations for the guiding
centers of the slow functional variables $w$ and $s$:
$$
\frac{dw}{dt} = - 2\gm_2 ( 1 - \al s) w + 2\gm_3 s^2 \; 
$$
\be
\label{68}
 \frac{ds}{dt} = - \gm_2  \al  w - \gm_3 s - \gm_1 ( s - s_\infty )  \;
\ee
in which $s_{\infty}$ is the equilibrium average spin and the effective interaction between
the sample and resonator is described by the coupling function
\be
\label{69}
 \al = g \; \frac{\gm^2}{\gm^2+\Dlt^2_s} \; ( 1 - As) 
\left \{ 1 - \left [ \cos(\Dlt_s t)\; - \; \frac{\Dlt_s}{\gm} \; \sin(\Dlt_s t)
\right ] e^{-\gm t} \right \} \;  .
\ee

According to the spin rotation symmetry of the system at the initial time, we impose
the zero initial conditions for $u(0) = 0$ and $w(0) = 0$. However, the spin polarization is
assumed to be non-zero and aligned along the static magnetic field, such that $s(0) = 1$.
Under these initial conditions, the system at~$t = 0$ is in a nonequilibrium state.
As soon as it starts at least slightly fluctuating due to spin waves, the feedback field
forces the total average spin to reverse aligning opposite to the static field $B_0$.
In~the process of the reversal, the transverse magnetization $u$ becomes non-zero,
which implies spin rotation symmetry breaking when
\be
\label{70}
  \lgl S_j^\pm \rgl \neq 0 \; .
\ee

The maximal absolute value of $u(t_0)$, occurring at time $t_0$, corresponds to the
maximal value of the coherence intensity $w(t_0)$. The latter depends also on the
detuning $w = w(t_0, \delta)$. Thus, the maximal spin rotation symmetry breaking
happens simultaneously with the maximal rotation coherence when
\be
\label{71}
 w(t_0,\dlt) = \max_t w(t,\dlt) \;  .
\ee

In this way, the effective order parameter can be defined as the normalized
maximal coherence~intensity
\be
\label{72}
\eta \equiv \frac{w(t_0,\dlt)}{w(t_0,0)} \;   .
\ee

From Equations (\ref{68}), we find the maximal coherence intensity
\be
\label{73}
 w(t_0,\dlt) = \left ( 1 \; - \; \frac{\gm\gm_2}{\gm_0\om_0} \; - \;
\frac{\gm_2\om_0}{\gm\gm_0}\; \dlt^2 \right )^2 \;  .
\ee

\begin{figure}[H]
\centering
\includegraphics[width=6cm]{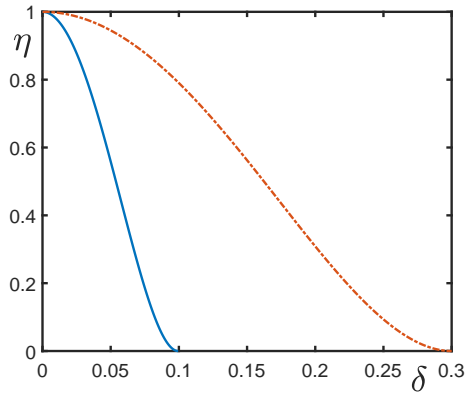}
\caption{Order parameter (\ref{75}) characterizing maximal coherence intensity
under spin-reversal resonance, with the parameters $\gamma_0/ \gamma_2 = 1$
and $\gamma/ \omega_0 = 0.1$ (dashed-dotted line) and $\gamma/ \omega_0 = 0.01$
(solid line).}
\label{f3}
\end{figure}

Thus, introducing the critical detuning
\be
\label{74}
 \dlt_c \equiv 
\sqrt{ \frac{\gm}{\om_0} \left ( \frac{\gm_0}{\gm_2} \; - \; \frac{\gm}{\om_0} \right ) } \;
\ee
we obtain the order parameter
\be
\label{75}
 \eta = \left ( 1 \; - \; \frac{\dlt^2}{\dlt_c^2} \right )^2 \;
\ee
whose behavior as a function of the relative detuning $\delta$ is demonstrated in Figure \ref{f3}.

It is worth noting that the non-zero transverse magnetization does not imply magnon
condensation but merely means that the average magnetic moment of the system
rotates around the $z$-axis, so that the total magnetization is not directed along this
axis \cite{Yukalov_40}. The correct introduction of magnons in a nonequilibrium picture
requires employing the Holstein--Primakov transformation with respect to the local in
time axis defined by the instantaneous time-dependent direction of the total average
magnetization \cite{Ruckriegel_41}.

\section{Conclusions}

We have demonstrated that resonance phenomena can be treated as a kind of
nonequilibrium phase transitions. Resonance
phenomena, similar to equilibrium phase transitions, are accompanied by symmetry breaking and can be described by order
parameters. Thus, under spin-wave resonance and helicon resonance, the
magnetization inside a metallic plate, induced by an incident electromagnetic field,
becomes periodic, with a slight perturbation caused by attenuation. In the case of
spin-reversal resonance, spin rotation symmetry breaks, and
the role of the order parameter is played by the coherence intensity. Experimental study of the behavior of the order parameters can yield information about the properties of the considered materials.

It is worth emphasizing that resonance phenomena are usually observed in finite systems.
Finite-width metallic plates were considered 
in the case of spin-wave and helicon resonances.
For the spin-reversal resonance, it was a finite sample that could be inserted in
a resonator electric coil of a finite volume. Symmetry breaking in finite systems is
a delicate topic, as it is quite different from the symmetry breaking in infinite systems
exhibiting equilibrium phase transitions. The effects of symmetry breaking in finite quantum
systems have recently been described in the review article \cite{Birman_42}. The
present paper can be considered as an additional chapter for this review.

\vskip 5mm

\authorcontributions{
The authors equally contributed to the paper.}

\vskip 5mm
{\parindent=0pt \small
{\bf Conflicts of Interest:} The authors declare no conflict of interest. }

\vskip 1cm


\end{document}